\documentclass[reqno,12pt]{amsart}
\newenvironment{dedication}
  {
   \thispagestyle{empty}
   \itshape             
   \raggedleft          
  }
  {\par 
  \vspace{10mm}
  }
\newtheorem{theo}{Theorem}
\newtheorem{rem}{Remark}

\newtheorem{prop}{Proposition}

\newtheorem{df}{Definition}

\newcommand\eps\varepsilon
\newcommand\ph\varphi
\newcommand\kap\Lambda

\usepackage{enumerate}
\usepackage{graphicx}
\usepackage{float}
\usepackage{yfonts}
\usepackage{mathrsfs}  
\usepackage{amsthm}
\usepackage{amsmath}
\usepackage{amscd}

\begin{document}\begin{dedication}
In the memory of \\
Professor Mikhail F. Sukhinin.\end{dedication}

\title[ Brachistochrone Problem and Vaconomic Mechanics]
{Brachistochrone Problem and Vaconomic Mechanics }


\author[Oleg Zubelevich]{Oleg Zubelevich\\ \\\tt
 Steklov Mathematical Institute of Russian Academy of Sciences\\
 \\Dept. of Theoretical mechanics,  \\
Mechanics and Mathematics Faculty,\\
M. V. Lomonosov Moscow State University\\
Russia, 119899, Moscow,  MGU \\ozubel@yandex.ru
 }
\email{ozubel@yandex.ru}
\date{}
\thanks{The research was funded by a grant from the Russian Science
Foundation (Project No. 19-71-30012)}

\begin{abstract}We consider different generalizations of the Brachistochrone Problem in the context of fundamental  concepts of  classical mechanics.
The correct statement for the Brachistochrone problem for nonholonomic systems is proposed.
It is shown that the Brachistochrone problem is closely related to  vakonomic mechanics.
\end{abstract}

\maketitle
\numberwithin{equation}{section}
\newtheorem{theorem}{Theorem}[section]
\newtheorem{lemma}[theorem]{Lemma}
\newtheorem{definition}{Definition}[section]

\section{Introduction. The Statement of the Problem}
The article is organized as follows. 

Section \ref{23wer} is independent on other text and contains an auxiliary material with precise definitions and proofs.
This section can be dropped by a reader versed in  the Calculus of Variations.

Section \ref{dfg690} is also pure mathematical. In this section we  justify the alternative statement of the variational problem for the  Brachistochrone with nonholonomic constraints. 

Other part of the text is less formal and based on Section \ref{23wer}. 

The Brachistochrone Problem is  one of the classical variational problems that we inherited form the past centuries. This problem was stated by  Johann Bernoulli in 1696 and 
  solved almost simultaneously  by him and by   Christiaan Huygens and Gottfried Wilhelm Leibniz.

Since that time the problem was discussed in different aspects numerous  times.

We  do not even try to touch this long and celebrated history.

This article is devoted to  comprehension of the  Brachistochrone Problem in terms of the modern Lagrangian formalism and to the generalizations which such a comprehension involves.

In the sequel we assume all the functions to be smooth. Recall that a function is smooth in the closed interval $[s_1,s_2]$ 
if by definition it belongs to $C^\infty(s_1,s_2)$ and all the derivatives are extended to continuous functions in $[s_1,s_2]$.

This assumption is overly strong  but we keep it for simplicity of the wording. 
\subsection{ Holonomic Version of the Problem}
Assume we are given with a Lagrangian mechanical system with a kinetic energy $T$ and a potential energy $V$:
\begin{equation}\label{56--}L=T-V,\quad T=\frac{1}{2}x_t^TG(x)x_t,\quad V=V(x).\end{equation}
By the subscript $t$ we denote the derivative in $t$.
Here $x=(x^1,\ldots,x^m)^T$ are the local coordinates on a configuration manifold $M$ and $G(x)$ is the matrix of a positively definite quadratic form, $$G^T(x)=G(x),\quad \det G(x)\ne 0,\quad\forall x\in M.$$ All the functions are smooth.

Fix two points $x_1,x_2\in M$. There are a lot of curves that connect these two points. Let a curve  $\gamma$ be one of them.  Assume that this curve is given by the equation \begin{equation}\label{xcv7;;}x=x(s),\quad x(s_1)=x_1,\quad x(s_2)=x_2.\end{equation}
So that $s$ is a coordinate in $\gamma$.

Impose an additional ideal constraint that makes  system (\ref{56--})  move along the curve $\gamma$ only.
We obtain a one-degree of freedom system with configuration space $\gamma$ and the generalized coordinate $s$. Motion of this system along the curve $\gamma$ is described  by the parameter $s$ such that $s=s(t)$ and 
\begin{equation}\label{dfg55}x=x(s(t)).\end{equation} 

Substituting this formula in (\ref{56--}) we obtain the Lagrangian of this new one-degree of freedom system:
$$L'(s,s_t)=\frac{1}{2}\big(x_s^T(s)G(x(s))x_s(s)\big)s^2_t-V(x(s)).$$
We want to choose  the curve $\gamma$  so that   system (\ref{56--}) spends minimal time  passing along $\gamma$  from  $x_1$ to $x_2$ at the energy level $h$:
\begin{equation}\label{569000}
\frac{1}{2}\big(x^T_s(s)G(x(s))x_s(s)\big)s^2_t+V(x(s))=h.\end{equation}
Let us assume that  
\begin{equation}\label{dfg4500}V(x)<h,\quad\forall x\in M.\end{equation}
Separating variables in (\ref{569000}) we see that the time of passing is given by the formula
$$\tau(\gamma)=\int_{s_1}^{s_2}\sqrt{\frac{x^T_s(s)G(x(s))x_s(s)}{2(h-V(x(s))}}ds.$$
Therefore we are looking for a stationary point of the functional $\tau$ under the boundary conditions (\ref{xcv7;;}).

By other words, the Brachistochrone curve is a geodesic of the Riemann metric
\begin{equation}\label{dfh678900}\frac{G(x)}{2(h-V(x))}.
\end{equation}

From proposition \ref{srgtsdgf} (see below) it follows that  we  can equivalently seek for a stationary point of the functional 
$$\mathfrak{T}(x(\cdot))=\int_{s_1}^{s_2}\frac{x^T_s(s)G(x(s))x_s(s)}{2(h-V(x(s)))}ds=\int_{s_1}^{s_2}\frac{T}{h-V}ds$$with the same boundary conditions. 

By the Hamilton principle it follows that the Brachistochrone curve is a trajectory of a dynamical system with Lagrangian 
$$\mathcal L(x,x_s)=\frac{T}{h-V}.$$

Introduce a function
$$W=-\frac{1}{h-V}.$$

Metric (\ref{dfh678900})  presents as follows
$$\frac{G(x)}{2(h-V(x))}=\frac{1}{2}(-W)G.$$
By the principle of least action in the Moupertuis-Euler-Lagrange-Jacobi form \cite{arn} we conclude that the Brachistochrone is a trajectory of a system with Lagrangian
$$\tilde L=T-W$$
at the zero energy level:
$$T+W=0.$$
\subsection{Holonomic Constraints}
Assume that the Brachistochrone problem is 
stated for the  system with Lagrangian (\ref{56--}) and with  ideal constraints
\begin{equation}\label{dfg45-=}
w(x)=0,\end{equation}
where $w=(w^1,\ldots w^n)^T(x),\quad n<m$ is a vector of smooth functions in $M$ such that
$$\mathrm{rang}\, \frac{\partial w}{\partial x}(x)=n,\quad \forall x\in M.$$
This statement brings nothing new: equations (\ref{dfg45-=}) define a smooth submanifold $N$ in $M$ and all constructed above theory works for the corresponding Lagrangian system  without constraints on $N$. 
\section{The Brachistochrone Problem with Differential Constraints}
\subsection{Discussion}
Assume that the Lagrangian system (\ref{56--}) is equipped with ideal  differential constraints
\begin{equation}\label{sfg2300-g}B(x) x_t=0.\end{equation}
Here 
$$B(x)=\begin{pmatrix}
b^1_1(x) & b_2^1(x) & \cdots & b_m^1(x) \\
b^2_1(x) & b^2_2(x) & \cdots & b^2_m(x) \\
\vdots  & \vdots  & \ddots & \vdots  \\
b^n_1(x) & b^n_2(x) & \cdots & b_m^n(x) 
\end{pmatrix}$$ is a matrix such that $$\mathrm{rang}\,B(x)= n<m,\quad \forall x\in M.$$
System (\ref{sfg2300-g}) can equivalently be expressed in terms of differential forms
$$\omega^i(x)=b^i_k(x)dx^k,\quad \mathcal T_x= \bigcap_{i=1}^n\ker \omega^i(x)\subset T_xM.$$
Namely,  $x_t(t)$ belongs to an element $\mathcal T_{x(t)}$ of the differential system $\{\mathcal T_x\}$
which is determined by equation (\ref{sfg2300-g}), $ \dim\mathcal T_x=m-n$.  

Recall that  if system (\ref{sfg2300-g}) can equivalently be presented in the form (\ref{dfg45-=}) then it is referred to as a holonomic system. Otherwise (\ref{sfg2300-g}) is a nonholonomic system. What  the word "equivalently" means and when such a presentation is possible is a content of the Frobenius theorem  \cite{shtern}.

As above we substitute equation (\ref{dfg55})   into Lagrangian (\ref{56--}) and separating the  variables in the energy integral (\ref{569000}) we get the functional $\tau$. Then we notice condition (\ref{dfg4500}). 

But now we must pick only motions that satisfy constraints (\ref{sfg2300-g}). Substituting (\ref{dfg55}) in (\ref{sfg2300-g}) we have
$$B(x(s(t))x_s(s(t))s_t=0.$$ It follows that
\begin{equation}\label{sdgwecc}B(x)x_s=0,\end{equation}
or equivalently $x_s(s)\in \mathcal T_{x(s)}$.

We postpone the discussion of the boundary conditions for a while but note that by the same reason 
(proposition \ref{srgtsdgf}) we can replace $\tau$ with $\mathfrak{T}$. 

We have obtained the problem of stationary points for the functional  $\mathfrak{T}$ defined on a set of curves $x=x(s)$ that obey constraints (\ref{sdgwecc}). Such a problem is called the Lagrange problem or the problem of vakonomic mechanics.

This problem provoked a lot of confusion and mistakes in classical mechanics. Many researches 
thought that such a variational problem is equivalent to the Lagrange-d'Alembert equations for 
the mechanical system with the Lagrangian $\mathcal L$   and ideal constraints (\ref{sdgwecc}) (the variable $s$ plays a role of time). 
If constrains (\ref{sdgwecc}) are nonholonomic it is not so.

Here is what in this concern Bloch, Baillieul, Crouch and Marsden write in \cite{bloh}:

{\it
It is interesting to compare
the dynamic nonholonomic equations, that is, the Lagrange-d’Alembert
equations with the corresponding variational nonholonomic equations. The
distinction between these two different systems of equations has a long and
distinguished history going back to the review article of Korteweg \cite{Korteweg [1899]}
and is discussed in a more modern context in Arnold, Kozlov, and Neishtadt \cite{akn}. (For Kozlov’s work on vakonomic systems see, e.g., \cite{Kozlov [1983]}
and \cite{Kozlov [1992]})

As
Korteweg points out, there were many confusions and mistakes in the literature
because people were using the incorrect equations, namely the variational equations,
when they should have been using the Lagrange-d’Alembert equations; some of these
misunderstandings persist, remarkably, to the present day. The upshot of the distinction is that the 
Lagrange-d’Alembert equations are the correct mechanical dynamical equations,
while the corresponding variational problem is asking a different question,
namely one of optimal control.

Perhaps it is surprising, at least at first, that these two procedures give
different equations. What, exactly, is the difference in the two procedures?
The distinction is one of whether the constraints are imposed before or
after taking variations. These two operations do not, in general, commute.
With the dynamic Lagrange-d’Alembert equations, we impose constraints
only on the variations, whereas in the variational problem we impose the
constraints on the velocity vectors of the class of allowable curves.}

In case of differential constraints (\ref{sdgwecc}) the situation with boundary conditions is much more complicated
than  (\ref{xcv7;;}). 

The main question is as follows.  Assume that the points $x_1,x_2\in M$ are connected with a  curve $x(s)$.
This curve is a stationary point of 
 $\tau$ or $\mathfrak{T}$  on the set of curves that satisfy constraints (\ref{sdgwecc}) and connect $x_1,x_2$.

Is the collection of other smooth paths that connect $x_1,x_2$ and satisfy (\ref{sdgwecc}) large enough to reduce the
variational problem to the differential equations? Or by other words, is this collection large enough to construct 
the  Lagrange multipliers method? In general the answer is "no". See remark \ref{srt438}  below. 

The author does not know whether the situation  will be fixed if we demand  the constraints to be completely nonholonomic.
Such questions seem to be  closely related to the Rashevsky-Chow theorem \cite{rash}, \cite{chow}. 

To avoid  these hard questions  
we suggest  considering the Brachistochrone Problem  with another boundary conditions which  guarantee the correct employment of the Lagrange multipliers method in the case of nonholonomic constraints.

\subsection{The boundary Conditions and the Lagrange multipliers method }

Let $\Sigma\subset M$ be a smooth $n-$dimensional submanifold such that
$$ x\in \Sigma\Longrightarrow T_xM=T_x\Sigma\oplus \mathcal T_x.$$

For  briefness of the story sake we proceed with local constructions. 

Any point $x\in \Sigma$ has a neighborhood $U\subset M$ such that  there are local coordinates $x$ in $U$:
\begin{equation}\label{wet45-0}x=\left(\begin{array}{c}y^1\\\vdots\\y^n\\z^1\\\vdots\\z^{m-n}\end{array}\right)=
\left(\begin{array}{c}y\\z\end{array}\right),\end{equation}
and
$$U\cap \Sigma=\{z=z_2\},\quad B(x)x_s=R(x)y_s+Q(x)z_s,$$
where $z_2$ is a  constant vector; $R$ is an $n\times n$ matrix and $$\det R(x)\ne 0,\quad \forall x\in U.$$

Let
us impose the boundary conditions:
\begin{equation}\label{jffmn9880}
x(s_1)=x_1=(y^T_1,z^T_1)^T\in U,\quad z(s_2)=z_2.\end{equation}
We  put no restrictions on $y(s_2).$

The geometric sense of these conditions is clear: the left end of the curve $x(s)$ is nailed at $x_1$ while the right one can slide along the surface $\Sigma:$
\begin{equation}\label{q1}x(s_1)=x_1,\quad x(s_2)\in\Sigma.\end{equation}

Observe that boundary conditions (\ref{jffmn9880}) differ from ones in the vakonomic mechanics \cite{akn}. In vakonomic mechanics the ends of the trajectory are fixed and the variations are not supposed to satisfy the equations of constraints. They satisfy the constraints in some asymptotic sense.

 \begin{theo}Let $\tilde x(s)$ be a stationary point of the functional $\mathfrak{T}$ on the set of functions $x(s)$ that satisfy
 (\ref{sdgwecc}), (\ref{jffmn9880}). 
 
 Then there is a smooth function $\lambda(s)=(\lambda_1,\ldots,\lambda_n)(s)$ such that $\tilde x$ satisfies the equations
\begin{equation}\label{xfg0900}\frac{d}{ds}\frac{\partial \mathcal L^*}{\partial x_s}-\frac{\partial \mathcal L^*}{\partial x}=0,\quad \mathcal L^*(s,x,x_s)=\mathcal L(x,x_s)+\lambda(s) B(x)x_s,\end{equation}
and
\begin{equation}\label{gtr4098}\frac{\partial \mathcal L}{\partial y_s}(\tilde x(s_2),\tilde x_s(s_2))+\lambda(s_2)R(\tilde x(s_2))=0.\end{equation}
\end{theo}
 This theorem is a direct consequence from theorem \ref{srt55678}.
 
 Condition (\ref{gtr4098}) can be presented in an invariant form:
 $$\frac{\partial \mathcal L^*}{\partial x_s}(s_2,\tilde x(s_2),\tilde x_s(s_2))v=0,\quad \forall v\in T_{\tilde x(s_2)}\Sigma.$$
 
 By proposition \ref{dfg67} the stationary point $\tilde x$ preserves the
 "energy":
 $$\frac{\tilde x^T_s(s)G(\tilde x(s))\tilde x_s(s)}{2(h-V(\tilde x(s)))}=\mathrm{const}.$$

Show that   system (\ref{xfg0900}), (\ref{sdgwecc})
can be presented in the normal form that is
\begin{equation}\label{vj776}x_{ss}=\Psi(x,x_s,\lambda),\quad \lambda_s=\Lambda(x,x_s,\lambda).\end{equation}
Indeed, system (\ref{xfg0900}) takes the form
\begin{equation}\label{sd5098}
  x_{ss}^T+\lambda_s\tilde B=\alpha(x,x_s,\lambda),\quad \tilde B=B\Big(\frac{G}{h-V}\Big)^{-1}.
\end{equation}
Differentiate (\ref{sdgwecc}) to have
\begin{equation}\label{sg099888}
B x_{ss}+\gamma(x,x_s)=0.\end{equation}
Substituting the second derivatives from (\ref{sd5098}) to (\ref{sg099888}) we obtain
\begin{equation}\label{60988}\lambda_s\tilde BB^T=\psi(x,x_s,\lambda).\end{equation}
It is clear  $\det \tilde BB^T\ne 0$ and we can express $\lambda_s$ from (\ref{60988}) and plug it in (\ref{sd5098}).

The way we construct (\ref{vj776}) shows that the vector-function
$$f(x,x_s)=B(x)x_s$$ is a first integral of (\ref{vj776}).

Observe also that equations (\ref{vj776}) are invariant under the substitution $$s\mapsto -s,\quad \lambda\mapsto -\lambda.$$

If we choose $y(s_2)$ and $z_s(s_2)$ then  $\lambda(s_2)$ and $y_s(s_2)$ are defined from equations  
(\ref{gtr4098}), (\ref{sdgwecc}); and at least for $s_2-s_1$ small, we can solve the Cauchy problem for (\ref{sd5098}), (\ref{60988}) backwards. 
Thus the problem is to find $y(s_2)$ and $z_s(s_2)$ so that the boundary conditions $x(s_1)=x_1$ are satisfied.

 These observations prompt the following theorem.
 
 \begin{theo}\label{sg1qwuuuu}Assume that $x_1$ is close enough to $\Sigma$ and $s_2-s_1>0$ is small enough. Then the problem
 (\ref {xfg0900}), (\ref{gtr4098}), (\ref{q1}), (\ref{sdgwecc}) has a unique smooth solution.\end{theo}
 \subsection{Proof of Theorem \ref{sg1qwuuuu}}
 Integrating  (\ref{vj776}) twice we get 
 \begin{equation}\label{sfg6ss}\begin{aligned}y_1&=(s_1-s_2)y_s(s_2)+y(s_2)+\int_{s_2}^{s_1}ds\int_{s_2}^sd\xi\, \Psi_y(x(\xi),x_s(\xi),\lambda(\xi)),\\
 z_1&=(s_1-s_2)z_s(s_2)+z_2+\int_{s_2}^{s_1}ds\int_{s_2}^sd\xi\, \Psi_z(x(\xi),x_s(\xi),\lambda(\xi)).
 \end{aligned}\end{equation}
 Here $\Psi=(\Psi_y^T,\Psi_z^T)^T.$
 
 Without loss of generality put $x_1=0$. Then the closeness of $x_1$ to $\Sigma$ means that $|z_2|$ is small.
 
 Taking into account  (\ref{sdgwecc}) rewrite (\ref{sfg6ss}) as
 \begin{equation}\label{poiuh}\begin{aligned}0&=-(s_1-s_2)R^{-1}(x(s_2))Q(x(s_2))z_s(s_2)+y(s_2)\\&+\int_{s_2}^{s_1}ds\int_{s_2}^sd\xi\, \Psi_y(x(\xi),x_s(\xi),\lambda(\xi)),\\
 0&=(s_1-s_2)z_s(s_2)+z_2+\int_{s_2}^{s_1}ds\int_{s_2}^sd\xi\, \Psi_z(x(\xi),x_s(\xi),\lambda(\xi)).\end{aligned}\end{equation}
 Introduce notations:
 $$\eps=s_2-s_1,\quad z_2=\eps \hat z_2,\quad y(s_2)=\eps\alpha_1,\quad z_s(s_2)=\alpha_2;$$
 and put
 $$C(\eps,\eps\alpha_1)=-R^{-1}(x)Q(x)\Big|_{x=\eps(\alpha_1^T,\hat z_2^T)^T}.$$
 System (\ref{poiuh}) takes the form
 \begin{equation}\label{sdrg00}
 \begin{aligned}F_1(\eps,\alpha_1,\alpha_2)&=-C(\eps,\eps\alpha_1)\alpha_2+\alpha_1+O(\eps)=0,\\
 F_2(\eps,\alpha_1,\alpha_2)&=-\alpha_2+\hat z_2+O(\eps)=0.\end{aligned}\end{equation}
 Here we little bit informally write  
 $$\int_{s_2}^{s_1}ds\int_{s_2}^sd\xi\, \Psi(x(\xi),x_s(\xi),\lambda(\xi))=O(\eps^2).$$
 
 The proof is accomplished by application of the Implicit Function Theorem to (\ref{sdrg00}):
 $$\det\Big(\frac{\partial F_i}{\partial \alpha_j}\Big)\ne 0,\quad \eps=0,\quad \alpha_2=\hat z_2,\quad \alpha_1=C(0,0)\hat z_2.$$

\section{Minimum of the Functional $\mathfrak{T}$}\label{dfg690}Here we prove that if $x_1$ is close enough to $\Sigma$ then the functional $\mathfrak{T}$ attains a minimum in the class of curves $x(s)$ such that 
(\ref{q1}) 
and  (\ref{sdgwecc}) holds. The uniqueness follows from theorem \ref{sg1qwuuuu}.

In this section all the inessential positive constants we denote  by  $c,c_1,c_2,\ldots$.

\subsection{The Main Spaces} For details on the Sobolev spaces see \cite{adams}.

Let $H^1(s_1,s_2)$ be the standard Sobolev space of vectors $$x(s)=(x^1,\ldots,x^m)^T(s)$$ such that
\begin{equation}\label{jhyy}\|x\|_{H^1(s_1,s_2)}^2=\sum_{i=1}^m\int_{s_1}^{s_2}\big((x^i(s))^2+(x_s^i(s))^2\big)ds<\infty.\end{equation}
Recall that the space $H^1(s_1,s_2)$ is compactly embedded in $C[s_1,s_2]$.

Introduce a subspace
$$H_0=\{x\in H^1(s_1,s_2)\mid x(s_1)=0\}.$$
A function
$$\|x\|^2=\sum_{i=1}^m\int_{s_1}^{s_2}\big(x_s^i(s)\big)^2ds$$ is a norm in $H_0$ and this norm is equivalent to the norm (\ref{jhyy}). Moreover,
\begin{equation}\label{f098uuu}\|x\|_{C[s_1,s_2]}\le c_1\sqrt{s_2-s_1}\|x\|.\end{equation}

Let a set $\mathcal J\subset H_0$ consist of functions $x(s)=(y^T,z^T)^T(s)$ such that
$$z(s_2)=z_2$$ and for almost all $s\in(s_1,s_2)$ equation (\ref{sdgwecc}) holds.

\subsection{Theorem of Existence of Minimum }Let $U\subset\mathbb{R}^m$ be a coordinate patch in $M$ with coordinates (\ref{wet45-0}). Let $|\cdot|$ stand for the standard Euclidean  norm in $\mathbb{R}^m$ and let $B_r(x)\subset \mathbb{R}^m$ stand for the open ball of radius $r$ and with center in $x$.

Take a   point $x'\in U\cap\Sigma$ such that $V(x')<h$. Then for some $\rho>0$ we have 
$$B_\rho(x')\subset U$$ and $\max\{V(x)\mid x\in\overline{B_{\rho}(x')}\}=V^*<h.$
Assume that the point $x_1$ is close to $x'$ such that $|x_1-x'|<\mu<\rho.$

  Without loss of generality put $x_1=0$. This particularly implies 
  \begin{equation}\label{sfg68} |z_2|<\mu.\end{equation}

Assume that for all $x\in M$ and for all $\xi\in \mathbb{R}^m$ we have
\begin{equation}\label{dfg20}\xi^T G(x)\xi\ge c_2|\xi|^2.\end{equation}
In addition to (\ref{dfg4500}) assume that \begin{equation}\label{dfg6ii}V(x)\ge c_9,\quad \forall x\in M.\end{equation}

\begin{theo} \label{sdg5600}

If $\mu>0$ is small enough then the functional $\mathfrak{T}$ attains its minimum in the set of functions
$$x:[s_1,s_2]\to U,\quad x(\cdot)\in \mathcal J.$$\end{theo}
It is not hard to show that the minimum $\tilde x$, as long as it exists, is actually smooth. Particularly $\tilde x$ satisfies equation (\ref{xfg0900}).

\subsection{Proof of Theorem \ref{sdg5600}}

Construct a function
$$ x_+(s)=(y^T_+, z^T_+)^T(s)\in \mathcal J$$ as follows
$$ z_+(s)=\frac{s-s_1}{s_2-s_1}z_2$$ and $ y_+$ is determined from the Cauchy problem
$$ (y_+)_s=A(s, y_+)\frac{z_2}{s_2-s_1},\quad  y_+(s_1)=0,$$
here
$$A(s,y)=-R^{-1}(x)Q(x)\Big|_{x=(y^T,z^T_+(s))^T}.$$
By the standard estimates we yield  
$$|y_+(s)|\le c_3\frac{s-s_1}{s_2-s_1}\mu,\quad s\in[s_1,s_2].$$
Here we use formula (\ref{sfg68}).

Thus $|x_+(s)|<c_4\mu$ and  for   $\mu>0$ small enough we obtain $$ x_+(s)\in B_\rho(x'),\quad \forall s\in[s_1,s_2],\quad x_+\in \mathcal J.$$ 
Observe that \begin{equation}\label{dfg60099}\mathfrak{T}(x_+)\le \frac{c_5\mu^2}{s_2-s_1}. \end{equation}

Let $\{u_k\}$ be a minimizing sequence for $\mathfrak{T}:$
$$\mathfrak{T}(u_k)\to\inf.$$
Then for all big enough $k$ it follows that
$$\mathfrak{T}(u_k)\le\mathfrak{T}(x_+).$$
Consequently, from formulas (\ref{dfg60099}), (\ref{dfg20}) and (\ref{f098uuu}) one concludes
$$\frac{c_7\|u_k\|^2_{C[s_1,s_2]}}{s_2-s_1}\le c_6\|u_k\|^2\le \frac{\mu^2}{s_2-s_1}.$$
This particularly implies
$$\|u_k\|_{C[s_1,s_2]}\le c_8\mu$$
and for   $\mu>0$ small enough we obtain $$ u_k(s)\in B_\rho(x'),\quad \forall s\in[s_1,s_2].$$ 

The further argument is completely standard   \cite{taylor}: since the sequence $\{u_k\}$ is bounded in $H^1(s_1,s_2)$ it  contains a subsequence $\{u_{k_i}\}$ that is convergent in $C[s_1,s_2]$ and weakly in $H^1(s_1,s_2)$. Therefore the exists an element
$\tilde u\in H_0$ such that
$$\|u_{k_i}-\tilde u\|_{C[s_1,s_2]}\to 0,\quad \int_{s_1}^{s_2}((u_{k_i})_s-\tilde u_s)^T\ph ds\to 0,\quad \forall\ph\in L^2(s_1,s_2).$$

The element $\tilde u$ is exactly the desired minimum  $\tilde x$.
This follows from the inequality:
\begin{align}\mathfrak{T}(u_{k_i})&\ge\mathfrak{T}(\tilde u)
+\frac{1}{2}\int_{s_1}^{s_2}\tilde u_s^T\Big(\frac{G(u_{k_i})}{h-V(u_{k_i})}-\frac{G(\tilde u)}{h-V(\tilde u)}\Big)\tilde u_sds\nonumber\\
&+\int_{s_1}^{s_2}\tilde u_s^T\Big(\frac{G(u_{k_i})}{h-V(u_{k_i})}-\frac{G(\tilde u)}{h-V(\tilde u)}\Big)\big((u_{k_i})_s-\tilde u_s\big)ds\nonumber\\
&+\int_{s_1}^{s_2}\tilde u_s^T\frac{G(\tilde u)}{h-V(\tilde u)}\big((u_{k_i})_s-\tilde u_s\big)ds.\nonumber\end{align}
The last three integrals vanish as ${k_i}\to\infty.$

The theorem is proved.

\section{Some Useful Facts From the Calculus of Variations}\label{23wer}
Here we collect several standard facts from the Calculus of Variations.

Let $\Omega\subset \mathbb{R}^m$ be an open domain with standard coordinates $$x=(x^1,\ldots,x^m)^T.$$ To proceed with formulations we split the vector $x$ in two parts as above (\ref{wet45-0}).

Let  $F:\Omega\times\mathbb{R}^m\to \mathbb{R}$ be a smooth function. 

We are about to state the variational problem for the functional
\begin{equation}\label{ste5}
\mathcal F\big(x(\cdot)\big)=\int_{s_1}^{s_2}F(x(s),x_s(s))ds\end{equation}
with 
boundary conditions
\begin{equation}\label{srt67}z(s_1)=  z_1,\quad z(s_2)=  z_2,\quad y(s_1)=  y_1,\quad s_1<s_2\end{equation}
and  constraints
\begin{equation}\label{dfg567}a(x, x_s)=0.\end{equation}
Here $a=(a^1,\ldots,a^n)^T$ is a vector of functions that are smooth in $\Omega\times \mathbb{R}^m$.

There also must be
$$(  y_1^T,  z_1^T)^T\in \Omega,\quad \{(y^T,z^T)^T\mid z=  z_2\}\cap \Omega\ne\emptyset.$$

Assume that 
\begin{equation}\label{sr43}
\det \frac{\partial a}{\partial y_s}(x,x_s)\ne 0,\quad (x,x_s)\in \Omega\times\mathbb{R}^m\end{equation}
and  equation (\ref{dfg567}) can equivalently be written as
$$
y_s=\Phi(y,z,z_s).$$

\begin{df}\label{dfg56}Let a smooth function 
$$\tilde x:[s_1,s_2]\to \Omega,\quad \tilde x(s)=(\tilde y^T,\tilde z^T)^T(s)$$
be such that
$$a(\tilde x(s),\tilde x_s(s))=0,\quad\tilde x(s_1)=  x_1=(  y_1^T,  z_1^T)^T,\quad \tilde z(s_2)=  z_2.$$
We shall say that $\tilde x$ is a  stationary point of  functional (\ref{ste5}) with constraints (\ref{dfg567}) and boundary conditions (\ref{srt67}) if the following holds.

For any smooth function $$X:[s_1,s_2]\times (-\eps_0,\eps_0)\to\mathbb{R}^m,\quad X(s,\eps)=(Y^T,Z^T)^T(s,\eps),\quad \eps_0>0$$
such that

1) $X\big([s_1,s_2]\times (-\eps_0,\eps_0)\big)\subset \Omega$;

2) $X(s,0)=\tilde x(s),\quad s\in [s_1,s_2]$;

3) $Z(s_1,\eps)=  z_1,\quad Z(s_2,\eps)=  z_2,\quad Y(s_1,\eps)=  y_1,\quad\eps\in(-\eps_0,\eps_0)$ ;

4) $a(X(s,\eps),X_s(s,\eps))=0,\quad (s,\eps)\in[s_1,s_2]\times (-\eps_0,\eps_0) $

we have 
$$\frac{d}{d\eps}\Big|_{\eps=0}\mathcal F\big(X(\cdot,\eps)\big)=0.$$

The functions $X$ with properties 1)-4) are referred to as variations.
\end{df}
\begin{theo}[\cite{ahiser}]\label{srt55678}
If the function $\tilde x$ is a  stationary point of  functional (\ref{ste5}) with constraints (\ref{dfg567}) and boundary conditions (\ref{srt67}) then there is a smooth function $\lambda(s)=(\lambda_1,\ldots,\lambda_n)(s)$ such that $\tilde x$ satisfies the equations
$$\frac{d}{ds}\frac{\partial F^*}{\partial x_s}-\frac{\partial F^*}{\partial x}=0,\quad F^*(s,x,x_s)=F(x,x_s)+\lambda(s) a(x,x_s),$$
and
\begin{equation}\label{s567}\frac{\partial F}{\partial y_s}(\tilde x(s_2),\tilde x_s(s_2))+\lambda(s_2)\frac{\partial a}{\partial y_s}(\tilde x(s_2),\tilde x_s(s_2))=0.\end{equation}
\end{theo}
This theorem remains valid if the functions $a,F$ depend on $s$.

 For completeness of the exposition sake we prove this theorem in  section \ref{xdfg56822}.
 
\subsection{The Homogeneous Case}In this section it is reasonable to assume the second argument of the functions $a,F$ to be defined on a conic domain $K\subset\mathbb{R}^m$. All the formulated above results and the argument of section   \ref{xdfg56822}  remain valid under such an assumption.

Recall that by definition the domain $K$ is a conic domain iff $$x\in K\Longrightarrow \alpha x\in K,\quad \forall\alpha>0.$$
 
 \begin{prop}\label{dfg67}
Assume that  the function $a$ is homogeneous in the second argument:
\begin{equation}\label{s3456-}a(x,\alpha  x_s)=\alpha a(a,x_s),\quad \forall\alpha>0,\quad \forall (x,x_s)\in \Omega\times K.\end{equation}
Then the stationary point $\tilde x$ preserves the "energy": $$H(x,x_s)=\frac{\partial F}{\partial x_s}x_s-F$$
 that is $H(\tilde x(s),\tilde x_s(s))=\mathrm{const}.$
  \end{prop}
{\it Proof of Proposition \ref{dfg67}.} Consider a function
$X(s,\eps)=\tilde x(s+\eps\ph(s))$ with a smooth function $\ph$ such that $\mathrm{supp}\,\ph\subset [s_1+s',s_2-s']$ 
and 
$$|\eps|,s'>0$$are small enough. 

The function $X$ satisfies all the conditions of Definition \ref{dfg56}. Property (\ref{s3456-}) is needed to check condition 4) of Definition \ref{dfg56}. 

Furthermore we have 
\begin{align}
 X&=\tilde x(s)+\eps\varphi(s) \tilde x_s(s)+O(\eps^2),\nonumber\\ X_s&=\tilde  x_s(s)+
\eps\big(\varphi_s(s) \tilde x_s(s)+\varphi(s) \tilde x_{ss}(s)\big)+O(\eps^2)\nonumber\end{align}
and 
\begin{align}
\frac{d}{d\eps}&\Big|_{\eps=0}\mathcal F( X(\cdot,\eps))\nonumber\\&=
\int_{s_1}^{s_2}\Big(\varphi(s)\frac{d}{ds}F(\tilde x,\tilde x_s)+\varphi_s(s)\frac{\partial F(\tilde x,\tilde x_s)}{\partial  x_s} \tilde x_s(s)\Big) ds\nonumber\\&
=\int_{s_1}^{s_2}H(\tilde x(s),\tilde x_s(s))\varphi_s(s) ds=0.\nonumber
\end{align}Here we use integration by parts.

Since $\ph$ is an arbitrary function the proposition is proved.

\subsubsection{The Both Functions $a,F$ are   Homogeneous in $x_s$}

\begin{prop}\label{sere32}Let $\tilde x(s)$ be  a stationary point of $\mathcal F.$
Define a function  $$\bar x(\xi)=\tilde x(f(\xi)),$$ where 
 $f:[\xi_1,\xi_2]\to[s_1,s_2]$ is a smooth function,
 $$f_\xi(\xi)>0,\quad f(\xi_r)=s_r,\quad r=1,2.$$

Then $\bar x$ is a stationary point of $\mathcal F$ with the integral taken over $[\xi_1,\xi_2].$ \end{prop}
Indeed, such a reparameterization neither changes the shape of constraints  (\ref{dfg567}) nor the shape of 
integral (\ref{ste5}):
\begin{align}\int_{s_1}^{s_2}F(X(s,\eps),X_s(s,\eps))ds&=\int_{\xi_1}^{\xi_2}F(\bar X(\xi,\eps),\bar X_\xi(\xi,\eps))d\xi, \nonumber\\\bar X(\xi,\eps)&=X(f(\xi),\eps).\nonumber\end{align}
\begin{prop}\label{sgf07}If $F(\tilde x(s),\tilde x_s(s))>0,\quad s\in[s_1,s_2]$ then for any constant $c>0$ we can choose a parametrization 
of $\tilde x$ such that 
$$F(\bar x(\xi),\bar x_\xi(\xi))=c,\quad\xi\in [\xi_1,\xi_2]. $$
\end{prop}
Indeed, the desired parametrization is obtained from the equation
$$\frac{d\xi}{ds}=\frac{1}{c}F(\tilde x(s),\tilde x_s(s)).$$

Introduce a functional 
$$\mathcal P\big(x(\cdot)\big)=\int_{s_1}^{s_2}\big(F(x(s),x_s(s))\big)^2ds$$ with the same constraints and boundary conditions (\ref{dfg567}), (\ref{srt67}).
\begin{prop}\label{srtertyuio}Let $\tilde x(s)$ be a stationary point of $\mathcal F$ and assume that $\tilde x(s)$ is parametrized in accordance with Proposition \ref{sgf07}:
$$F(\tilde x(s),\tilde x_s(s))=c.$$
 Then $\tilde x(s)$ is a stationary point of $\mathcal P$.\end{prop}
 Indeed,
 \begin{equation}\label{fdsssss}\frac{d}{d\eps}\Big|_{\eps=0}\mathcal P\big(X(\cdot,\eps)\big)=2c\frac{d}{d\eps}\Big|_{\eps=0}\mathcal F \big(X(\cdot,\eps)\big)=0.\end{equation}
 \begin{prop}\label{sf4yuitt} Let $x^*(s)$ be a stationary point of $\mathcal P$. Then $F$ is the energy integral:
 $$F(x^*(s),x^*_s(s))=\mathrm{const}.$$\end{prop}
 Indeed, since $F$ is  homogeneous in $x_s$
it follows that 
$$\frac{\partial F}{\partial x_s}x_s=F.$$ And the assertion follows from Proposition \ref{dfg67}:
$$H=\frac{\partial F^2}{\partial x_s}x_s-F^2=F^2.$$
\begin{prop}\label{sfwp}
 Let $x^*(s)$ be a stationary point of $\mathcal P$ such that
 $$F(x^*(s),x^*_s(s))=c>0.$$ Then it is a stationary point of $\mathcal F$.\end{prop}
 Indeed, it follows from (\ref{fdsssss}).
 
 Summing up we obtain the following proposition.
 \begin{prop}\label{srgtsdgf}If $\tilde x(s)$ is a stationary point of $\mathcal F$ 
 and $F(\tilde x(s),\tilde x_s(s))>0$ then after some reparametrization it is a stationary point of $\mathcal P$.
 
 If $ x^*(s)$ is a stationary point of $\mathcal P$ and $F( x^*(s), x^*_s(s))=c>0$ then it is a stationary point of $\mathcal F$.\end{prop}

\subsection{Proof of Theorem \ref{srt55678}}\label{xdfg56822}
Introduce a notation
$$[F]_y=-\frac{d}{ds}\frac{\partial F}{\partial y_s}+\frac{\partial F}{\partial y},\quad
[F]_z=-\frac{d}{ds}\frac{\partial F}{\partial z_s}+\frac{\partial F}{\partial z}$$
and correspondingly $[F]_x=([F]_y,[F]_z)$.

Let us put $Z(s,\eps)=\tilde z(s)+\eps\delta z(s),$
\begin{equation}\label{sef321} \mathrm{supp}\,\delta z\subset[s_1,s_2].\end{equation}
Then the function $Y$ is uniquely determined from the following Cauchy problem
\begin{equation}\label{fgzz55}Y_s(s,\eps)=\Phi(Y(s,\eps),Z(s,\eps),Z_s(s,\eps)),\quad Y(s_1,\eps)=  y_1.\end{equation}
\begin{rem}\label{srt438}That is why we can not impose condition $ x(s_2)=  x_2$  as 
it is usually done for the holonomic case. The value $Y(s_2,\eps)$ has already been uniquely defined  by other boundary conditions and the constraints. In other words if we add the  condition $Y(s_2,\eps)=  y_2$ to the conditions 1)-4) of  Definition \ref{dfg56} then the set of variations $\{X(s,\eps)\}$ may turn up to be insufficiently large to prove theorem \ref{srt55678}.

For example, consider a plane $\mathbb{R}^2=\{x=(y,z)^T\}$. There is a unique smooth path from $x_1=(0,0)^T$ to $x_2=(0,1)^T$
that  satisfies the equation $y_s=0$. (Much more complicated example by C. Caratheodory see in \cite{akn}.)
\end{rem}
 Cauchy problem (\ref{fgzz55}) has the suitable solution at least for $|\eps|$ and $\,s_2-s_1$ small. Observe also that
\begin{equation}\label{sdrg44}
Y_\eps(s_1,\eps)=0.\end{equation}
Using the standard integration by parts technique and from formulas (\ref{sdrg44}), (\ref{sef321}) we obtain
\begin{align}
\frac{d}{d\eps}\Big|_{\eps=0}&\mathcal F\big(X(\cdot,\eps)\big)\nonumber\\
&=\int_{s_1}^{s_2}\Big([F]_z\delta z+[F]_yY_\eps\Big)ds+\frac{\partial F}{\partial y_s}(\tilde x(s_2),\tilde x_s(s_2))Y_\eps(s_2,0)=0.\label{wer11}\end{align}
The function $\lambda(s)$ is still undefined but  due to condition (\ref{sr43}) the value $\lambda(s_2)$ is determined uniquely from (\ref{s567}).

From condition 4) of definition \ref{dfg56} it follows that
$$A(\eps)=\int_{s_1}^{s_2}\lambda(s)a(X(s,\eps),X_s(s,\eps))ds=0.$$
By the same argument as above we have
\begin{align}
\frac{d}{d\eps}&\Big|_{\eps=0}A=\int_{s_1}^{s_2}\Big([\lambda a]_z\delta z+[\lambda a]_yY_\eps\Big)ds\nonumber\\&+\lambda(s_2)\frac{\partial a}{\partial y_s}(\tilde x(s_2),\tilde x_s(s_2))Y_\eps(s_2,0)=0.\label{w11}\end{align}
Summing formulas (\ref{w11}) and (\ref{wer11}) we yield 
\begin{equation}\label{sdf2309899}\int_{s_1}^{s_2}\Big([F^*]_z\delta z+[F^*]_yY_\eps\Big)ds=0.\end{equation}
To construct the function $\lambda$ consider an equation 
\begin{equation}\label{drg1234}[F^*]_y=0.\end{equation} This is a system of linear ordinary differential equations for $\lambda$. Due to assumption  (\ref{sr43}) this system can be presented in the normal form that is
$$\lambda_s=\Lambda(s,\lambda).$$
Since we know $\lambda(s_2)$, by the existence and uniqueness theorem we obtain $\lambda(s)$ as a solution to the IVP for (\ref{drg1234}). 

Equation (\ref{sdf2309899}) takes the form 
$$\int_{s_1}^{s_2}[F^*]_z\delta zds=0.$$
Since $\delta z$ is an arbitrary function we get $[F^*]_z=0$. Together with (\ref{drg1234}) this proves the theorem.

\section{Appendix: Generalizations and the Hamiltonian point of view}
This section is independent on the previous text and contains a method of reduction of the vaconomic mechanics problems to the Hamilton equations.

Let $M$ be a smooth $m-$dimensional manifold with local coordinates $x=(x^1,\ldots,x^m)^T$. We assume all the objects below  to be  smooth enough for the formulas to make sense. 

Let $L:TM\to \mathbb{R},\quad L=L(x,v)$ be a Lagrangian function. Assume that the mapping
\begin{equation}\label{xfg6}(x,v)\mapsto (x,y),\quad y=\frac{\partial L}{\partial v}(x,v)\end{equation} is a diffeomorphism of $TM$ to $T^*M$ and equation (\ref{xfg6}) implies
$$v=f(x,y),\quad f=(f^1,\ldots,f^m)^T.$$Here $(x,y)$ are the local coordinates in $T^*M$.

Assume also that  $$A(x,v)=\frac{\partial^2 L}{\partial v^2}(x,v)$$ is a matrix of  positively definite quadratic form;
\begin{equation}\label{dfg66}\frac{\partial f}{\partial y}=A^{-1}.\end{equation}

Introduce $1-$forms
$$b^i=b^i_j(x)dx^j,\quad i=1,\ldots,n<m;$$ and
$$B(x)=\begin{pmatrix}
b^1_1(x) & b_2^1(x) & \cdots & b_m^1(x) \\
b^2_1(x) & b^2_2(x) & \cdots & b^2_m(x) \\
\vdots  & \vdots  & \ddots & \vdots  \\
b^n_1(x) & b^n_2(x) & \cdots & b_m^n(x) 
\end{pmatrix},\quad \mathrm{rang}\,B(x)= n,\quad  x\in M.$$ 
These forms define an $m-n-$dimensional differential system
$$\mathcal T_x=\bigcap_{i=1}^n\ker b^i(x)\subset T_xM$$
in $M$ \cite{shtern}.

Let $\Sigma\subset M$ be a smooth $n-$dimensional submanifold such that
$$ x\in \Sigma\Longrightarrow T_xM=T_x\Sigma\oplus \mathcal T_x.$$

Let $S$ stand for a  set of functions $x:[t_1,t_2]\to M$ such that
$$x(t_1)=x_1,\quad x(t_2)\in\Sigma,\quad \dot x(t)\in\mathcal T_{x(t)},\quad t\in[t_1,t_2].$$ Here $x_1\in M$
is a fixed point.

In coordinate notation the last inclusion is read as follows
\begin{equation}\label{sdfg66}
B\big(x(t)\big)\dot x(t)=0.\end{equation}
The  scheme described below works if we replace constraints (\ref{sdfg66}) with   ones nonlinear in $\dot x$. We use linear constraints just to
avoid an  irrelevant   complication of the technique.

From the calculus of variations \cite{ahiser} we know that  if a function $\tilde x(\cdot)\in S$ is a critical point of the functional
$$F:S\to\mathbb{R},\quad F\big(x(\cdot)\big)=\int_{t_1}^{t_2}L\big(x(t),\dot x(t)\big)dt$$ then
there is a vector $\lambda(t)=(\lambda_1,\ldots,\lambda_n)(t)$ such that the function $\tilde x(t)$ satisfies the Lagrange equations
\begin{equation}\label{dfboop}
\frac{d}{dt}\frac{\partial L^*}{\partial v}\big(x,\dot x,\lambda(t)\big)-
\frac{\partial L^*}{\partial  x}\big(x,\dot x,\lambda(t)\big)=0,\end{equation}
with $$
 L^*(x,v,\lambda)=L(x,v)+\lambda B(x)v,\quad L^*:TM\times\mathbb{R}^n\to\mathbb{R}.$$

The main object of our study is system (\ref{sdfg66}),  (\ref{dfboop}). 
Loosely speaking our aim is to show that this system is described in terms of a Hamiltonian system in some $2m-$dimensional symplectic manifold.

As far as the author knows this simple but important  fact has not been noticed anywhere.

\subsection{The Main Construction}Let $$H^*:Q\to\mathbb{R},\quad Q=T^*M\times \mathbb{R}^n $$ stand for the standard Legendre transformation of $L^*$:
\begin{align}p&=(p_1,\ldots,p_m)=\frac{\partial L^*}{\partial v}(x,v,\lambda),\quad v=\tilde f(x,p,\lambda),\quad\mbox{where}\nonumber\\
\tilde f(x,p,\lambda)&:=f\big(x,p-\lambda B(x)\big);\nonumber\\
H^*(x,p,\lambda)&=p\tilde f(x,p,\lambda)-L\big(x,\tilde f(x,p,\lambda)\big)-\lambda B(x)\tilde f(x,p,\lambda).\nonumber\end{align}

The following well-known fact holds true. Let the function $\lambda=\lambda(t)$ be fixed then the function $x(t)$ is a solution to equations (\ref{dfboop}) iff the functions 
$$x(t),\quad p(t)=\frac{\partial L^*}{\partial \dot x}\big(x(t),\dot x(t),\lambda(t)\big)$$ are the solution to the Hamilton equations
\begin{equation}\label{sg55}\dot x=\frac{\partial H^*}{\partial p}\big(x,p,\lambda(t)\big),\quad \dot p=-\frac{\partial H^*}{\partial x}\big(x,p,\lambda(t)\big).\end{equation}
Equation (\ref{sdfg66}) takes the form
\begin{equation}\label{srg00}
B(x)\tilde f(x,p,\lambda)=0.\end{equation}
To determine the function  $\lambda(t)$ one must differentiate this equation in $t$ and substitute $\dot x,\dot p$ from (\ref{sg55}). This procedure gives an equation of the form
$$R(x,p,\lambda)\dot\lambda^T=\psi(x,p,\lambda),\quad R(x,p,\lambda)=B(x)\frac{\partial f}{\partial y}\big(x,p-\lambda B(x)\big)B^T(x).$$
Recall that   $A$ is a matrix of positive definite quadratic form. So that by formula (\ref{dfg66}) we have 
\begin{equation}\label{09}\det R(x,p,\lambda)\ne 0;\end{equation} and the last equation takes the form
\begin{equation}\label{sg000}\dot\lambda=\Lambda(x,p,\lambda).\end{equation}

System (\ref{sg55}), (\ref{sg000}) is an ODE system on the manifold $Q$ with the local coordinates $(x,p,\lambda).$

Observe that the way we construct  system (\ref{sg55}), (\ref{sg000}) implies that the vector function
$$C(x,p,\lambda)=B(x)\tilde f(x,p,\lambda)$$ is a first integral of this system. This in particular means that a 
solution $(x,p,\lambda)(t)$ to system (\ref{sg55}), (\ref{sg000}) satisfies (\ref{srg00}) identically provided the solution's  initial conditions satisfy (\ref{srg00}).

Let $N$ denote the zero level manifold of $C(x)$:
$$N=\{(x,p,\lambda)\in Q\mid C(x,p,\lambda)=0\}.$$
\begin{prop}\label{90jju}The set $N$ is a smooth $2m-$dimensional submanifold of $Q$; this submanifold is locally presented as a graph of a function:
\begin{equation}\label{dfb098}\lambda=\ph(x,p).\end{equation}
\end{prop}
Indeed, this follows from the Implicit Function Theorem and formula (\ref{09}) : 
$$\frac{\partial C}{\partial \lambda}=R.$$

\begin{rem}\label{xfg000l)}Formula (\ref{dfb098}) implies that $(x,p)$ are the local coordinates in $N$.\end{rem}

The manifold $T^*M$ has the standard symplectic structure:  $$\Omega=dx^i\wedge dp_i.$$ This $2-$form is trivially extended to the cross product $Q$. Therefore we proceed with considering of $\Omega$ in $Q$.

\begin{prop}The manifold $N$ is a symplectic manifold with a symplectic form given by the following restriction
$$\omega=\Omega\mid_{N}.$$
The coordinates $(x,p)$ are symplectic coordinates in $N$.
\end{prop}
Indeed, this is a direct consequence of proposition \ref{90jju} and remark \ref{xfg000l)}.

Let $$v(z)\in T_zQ,\quad z=(x,p,\lambda)$$ denote the vector field of  system (\ref{sg55}), (\ref{sg000}). 
Since $C$ is a first integral we have $z\in N\Rightarrow v(z)\in T_zN$.

\begin{theo}\label{sdg00hh}The vector field $u=v\mid_N$  is a Hamiltonian vector field in $N$. The  Hamiltonian of $u$ is $$H=H^*\mid_N:N\to\mathbb{R}.$$
In the local coordinates $(x,p)$ this Hamiltonian is given by the formula
$$H(x,p)=H^*\big(x,p,\ph(x,p)\big).$$\end{theo}

{\it Proof of Theorem \ref{sdg00hh}.}
Let $i_v$ stand for the  interior product. The following formula is the main:
\begin{equation}\label{sd00}i_v\Omega=dH^*-\frac{\partial H^*}{\partial\lambda_j}d\lambda_j.\end{equation}
This formula is checked by direct calculation.

The formulas
$$\frac{\partial H^*}{\partial\lambda}=-B(x)\tilde f(x,p,\lambda),\quad \frac{\partial H^*}{\partial\lambda}\Big|_N=0$$
are evident.

Restricting the both sides of formula (\ref{sd00}) to $N$ we obtain
$$i_u\omega=dH.$$

The theorem is proved.

\subsection{Example}Let the Lagrangian $L$ be given by the formula
$$L(x,\dot x)=\frac{1}{2}\dot x^TG(x)\dot x,$$ where $G(x)$ is a Riemann metric in $M$.

After some calculation one yields
$$H(x,p)=\frac{1}{2}p\mathscr P(x)G^{-1}(x)p^T,$$
where
\begin{align}\mathscr P(x)=E&-G^{-1}(x)B^T(x)\big(B(x)G^{-1}(x)B^T(x)\big)^{-1}B(x),\nonumber\\
\mathscr P(x)&:T_xM\to T_xM.\nonumber\end{align} is the orthogonal projection:
$$\mathscr P(x) T_xM=\mathcal T_x.$$

\subsection*{Acknowledgments} The author wishes to thank Professor V. V. Kozlov for useful discussions.

\end{document}